\documentclass[a4paper,11pt]{article}
\pdfoutput=1 
\usepackage[latin1]{inputenc}
\usepackage[T1]{fontenc} % if needed
\usepackage{graphicx, multirow}
\usepackage{subcaption}
\usepackage{float}
\usepackage{bm}
\usepackage{braket}
\usepackage{amsmath}
\usepackage{amssymb}
\usepackage{amscd}
\usepackage{booktabs}
\usepackage{latexsym}
\usepackage{slashed}
\usepackage{color, xcolor}
\usepackage{graphicx}
\usepackage[normalem]{ulem}
\definecolor{orange}{cmyk}{0,0.5,1,0}
\usepackage{verbatim}
\usepackage{rotating}
\usepackage{cancel}
\usepackage{caption}
\usepackage{hyperref}
\usepackage{lineno}
\usepackage{makecell,cite}
\hypersetup{
   colorlinks=true,     % false: boxed links; true: colored links
   linkcolor=blue,      % color of internal links
   citecolor=red,       % color of links to bibliography
   filecolor=magenta,   % color of file links
   }
\newcommand{\beq}{\begin{equation}}
\newcommand{\eeq}{\end{equation}}
\newcommand{\bea}{\begin{eqnarray}}
\newcommand{\eea}{\end{eqnarray}}

\begin{document}
%\title{}  
%\author{}
%\affiliation{}
%\author{}
%\affiliation{} 
%\author{}
%\affiliation{} 
%---------------------------------------
\begin{center}

{\large \bf Probes of Anomalous Events at LHC with Self-Organizing Maps} \\    
\vskip 0.6cm
Shreecheta Chowdhury$^{a}$%\footnote{shreechetac@srmap.edu.in}
, 
Amit Chakraborty$^{a}$%\footnote{aamit.phy@gmail.com} 
,  
and
Saunak Dutta$^{b}$%\footnote{* On leave}
\vskip 0.6cm
{$^a$  Department of Physics, SRM University-AP, Amaravati, Mangalagiri 522240, India}
\vskip 0.1cm
{$^b$ School of Science and Technology, Vijaybhoomi University, Karjat, Raigad 410201, India.} \\

\end{center}
\vskip 0.3cm
%%%%%%%%%%%%%%%%%%%%%%%%%%%%%%%%%%%%%%%%
%\maketitle
%\linenumbers 

%%-------------------- Document begins --------------------%%
%\newpage
%\setcounter{footnote}{0}
%%%%%%%%%%%%%%%%%%%%%%%%%%%%%%%%%%%%%%%%%%%%%%%
\begin{abstract}

We propose an Unsupervised Learning Algorithm, Self-Organizing Maps (SOM), built on a neural network architecture, for the probe of a rare top decay, mediated by Flavor Changing Neutral Current (FCNC), to charm and the Higgs boson, with the Higgs boson further decaying to a pair of b-quarks or a pair of gauge bosons ($W^{\pm}/Z$) in a boosted regime. Ideally, the particles originating from the decay of the boosted top lead to the reconstruction of a large-R jet, comprising three-prong substructures, with b- and c-tagged subjets.
{{The SOM algorithm has been demonstrated as a model-agnostic anomaly-finder for probing the rare decay at the LHC, by mapping distinct signal and background regions to separate non-overlapping clusters on the Kohnen map. This helps to identify signal regions with higher signal significances. We also discuss the robustness of this algorithm, especially for other BSM probes with model-agnostic and model-dependent searches.}}

%{\textcolor{blue}{The SOM algorithm has been demonstrated as an effective model-agnostic anomaly-finder for probing the rare top decay at the LHC, by mapping distinct signal and background regions to separate non-overlapping clusters on the Kohnen map. This helps to understand the data better and identify signal regions with higher signal significances.}} %We have illustrated that SOM can be effectively implemented as an anomaly-finder for model-agnostic probes of such decay at the LHC. Relatively simple in formalism, this algorithm achieves commendable performance over other existing formalisms, tailored to probe this rare decay. 
%We also discuss the robustness of this algorithm, which can be successfully implemented for other BSM probes, both for model-agnostic and model-dependent searches. 

\end{abstract}

%\newpage
\vskip 1.5cm
%%%%%%%%%%%%%%%%%%%%%% Table of content %%%%%%%
\hrule
\tableofcontents
\vskip 1.0cm
\hrule

%%%%%%%%%%%%%%%%%%%%%%%%%%%%%%%%%%%%%%%%%%%%%%%
\newpage

\section{Introduction}  \label{intro}

The Standard Model of Particle Physics (SM) has been an accomplished framework encapsulating strong and weak interactions of nature \cite{Sakurai1960, Scheck:1996ur, Pich:2007vu, Pich:2005mk}. Albeit its success, theoretical inconsistencies, and observational discrepancies motivated theories Beyond the Standard Model (BSM) \cite{Giudice2017,Sakharov1967,Peccei1977}. With several BSM theories proposed \cite{2010arXiv1005.1676L}, and with recent upgrades at the Large Hadron Collider (LHC) underway \cite{Apollinari:2015bam, Giovannozzi:2013pwa}, the identification of events governed by a specific BSM formalism over abundant SM backgrounds is becoming increasingly challenging, establishing the pursuit of advanced and effective identification strategies is a necessity. 

Recent advancements in Machine Learning (ML) algorithms have progressively replaced traditional cut-based search techniques to identify boosted objects at the LHC \cite{Larkoski:2017jix,Radovic:2018dip,Plehn:2022ftl,MachineDAP,Feickert:2021ajf}. According to the ML dialect, BSM searches with supervised learning techniques reduce to a classification problem for a model-dependent probe, where events are categorized individually as signals or backgrounds depending on their respective traits. On the contrary, model-agnostic probes with unsupervised learning techniques identify anomalies over patterns anticipated from collisions governed by the SM framework on a collective dataset. In this paper, we introduce an ML algorithm, the Self-Organizing Maps (SOM) \cite{Kohonen_1982} and demonstrate its effectiveness for the latter search strategy.

The landscape of ML-driven top-tagging has evolved significantly. Traditional methods \cite{Kaplan:2008ie, Plehn:2010st} have been surpassed by sophisticated approaches, encompassing both model-dependent and model-agnostic probes leveraging respectively, supervised and unsupervised learning strategies. For a glimpse into the diverse array of such topics, we refer the reader to \cite{Butter:2017cot,Macaluso:2018tck,Chakraborty:2020yfc,Bhattacherjee:2022gjq,Furuichi:2023vdx,Brehmer:2024yqw,Dong:2024xsg,Choi:2023slq,
Collins:2018epr,Amram:2020ykb,Cheng:2020dal,Pol:2020weg,Finke:2021sdf,Ostdiek:2021bem,Ngairangbam:2021yma,Alvi:2022fkk,Dillon:2022tmm,Golling:2023yjq,Bickendorf:2023nej,Metodiev:2023izu,Hammad:2024dsn}, and \cite{10.21468/SciPostPhys.7.1.014, Bose:2024pwc} for a comprehensive review of the domain. 
However, there has been limited exploration of SOM \cite{Habermann:2023ksr} in Particle Physics Phenomenology. The scope of this work is a portrayal of SOM as an anomaly-probing prescription for model-independent searches, demonstrated over FCNC-mediated rare decay of top quark at LHC.

Self-Organizing Map resembles a planar grid, with nodes placed over the vertices. Each node represents a specific event topology. During training, SOM organizes events with similar traits together using an iterative approach, forming groups that capture events with underlying similarities together. In this paper, we explore the potential of this algorithm as an effective tool for anomaly detection, as well as an efficient identifier of a specific BSM signature. We observe that SOM possesses promising implications, associated not only with the identification of the rare decay processes over an overwhelming SM background but also with the segregation of different decay modes of heavy states with similar event topologies. 

As an illustration, we probed a rare top decay, driven by a flavor changing neutral current (FCNC), to a c-quark and Higgs boson $\left( t \to c H \right)$ with suppressed branching fraction $\lesssim 0.1\%$ \cite{CMS:2024ubt, CMS:2021gfa} in a boosted regime ($p^{top}_T >$ 350 GeV), over the background dominated by the most abundant decay mode of the top $\left( t \to b W \right)$, and spurious QCD events $\left( p p \to j j \right)$ mimicking the signal topology. The decay products of the boosted top are highly collimated, enveloped by a large-R jet comprising a three-prong substructure at LHC. Kinematic features specific to the signal $\left( t \to c H \right)$, irreducible background $\left( t \to b W \right)$, and the QCD di-jet background, are mapped on the planar grid which, through iterative adjustment of weights attributed on the nodes, leads to the identification of individual decay modes as clusters appearing on the grids. {By clustering the input data, the SOM algorithm effectively guided the determination of optimal signal regions, achieving a signal significance of $\sim 2 \sigma$ for $\mathcal{BR}( t \to cH)$ = 0.05\% at 3 ab$^{-1}$ over SM backgrounds, for probing the rare top decay mediated by FCNC.}
This is a fair leap as compared to our previous endeavor \cite{Chowdhury:2023jof}, where we presented a simple ML model for the same probe that outperformed existing top-tagging methodologies \cite{10.21468/SciPostPhys.7.1.014}, tailored to probe the said decay. %\textcolor{red}{ This probe of rare top decay mediated by FCNC using SOM achieves a signal significance of $\sim 2 \sigma$ for $\mathcal{BR}( t \to cH)$ = 0.05\% at 3 ab$^{-1}$ over SM backgrounds, a fair leap as compared to our previous endeavor \cite{Chowdhury:2023jof}, where we presented a simple ML model for the same probe that outperformed existing top-tagging methodologies \cite{10.21468/SciPostPhys.7.1.014}, tailored to probe the said decay.} 
Finally, we provide an interpretation for each cluster boundary formed on the SOM grid, drawing insights into the relevant features underlying the event topologies of the three different processes under consideration.

The approach followed for this work is model-independent and robust; similar prescriptions can be formulated for BSM probes involving boosted composite topologies. With anticipated operations of future particle colliders in higher energy frontiers, we find the analysis of boosted objects pertinent, with potentials for witnessing New Physics breakthroughs in the near future. 

This paper has been organized as follows: Section \ref{algo} provides an overview of the SOM algorithm, Section \ref{Event_gen} outlines the event generation, object reconstruction, and input feature selection processes,  Section \ref{Result} discusses the results obtained from the training, providing insights into the patterns reflected on the trained weight vectors on the grid, Section \ref{BSM} articulates the optimal signal region for the FCNC top decay process ($t \to c H$), and Section \ref{summary} provides the concluding remarks, future directives, and highlights of the entire exploration.

%%%%%%%%%%%%%%%%%%%%%%%%%%%%%%%%%%%%%%%%%%%%%%%%%%%

\section{Self Organizing Maps: The Algorithm }  \label{algo}

The Self Organizing Maps (SOM) is an Unsupervised Machine Learning algorithm that generates a lower-dimensional representation of a higher-dimensional dataset, simultaneously preserving the topological structure of the data. Through Competitive Learning, SOM learns to pin down the position of every training data-point (simulated events in this context) on the topographic map of neurons. Consequently, dimensionality reduction makes SOM a perfect tool for data visualization. 

An SOM, unlike generic Deep Learning architectures, does not require a target output for training. Instead, where the weights assigned to a given node match the input vector (a specific event), the weights on the nodes present in the neighborhood of the Best Matching Unit (BMU) are selectively optimized to resemble the data point, subsequently capturing the underlying traits of the class (signal or different backgrounds) the input vector belongs to. 

From an initial distribution of random weights and over subsequent iterations, SOM eventually converges into a map of stable zones, through an iterative process of determining the node best resembling a given input event, labeled the Best Matching Unit (BMU), thereby updating the weights of neurons locally encircling the BMU, in accordance with the neighborhood function and respective differences from the input data. 

The updated weight vector for a specific node is given by: 
\bea 
W_v (t+1) = W_v(t) + \eta(t) \cdot \phi(u,v,t) \cdot [(X(t) - W_v (t))],
\label{Eqn:WeightUpdates}
\eea
where, $v$ are the coordinates of the said node, $u$ are the coordinates of the BMU, $W_v(t)$ is the weight vector of the said node at time step $t$, $X(t)$ is the input {event} presented at time step $t$, $\eta(t)$ is time-dependent learning rate, and $\phi(u,v,t)$ is the Neighborhood Distance Function for the BMU (at $u$) and the concerned node (at $v$) at the time step $t$. The learning rate $\eta(t)$ follows an exponential damping given as, 
\bea 
\eta(t) = \eta(0) \cdot \exp{(-t/\lambda)},
\label{Eqn:LRate}
\eea
with $\eta(0)$ being the initial learning rate and $\lambda$ being the learning decay rate. The Neighborhood Distance Function follows a Gaussian distribution pattern. It decays rapidly with the distance between the said node and the BMU, with the decay rate impeded by a time-dependent factor mimicking the variance of the Gaussian distribution. Hence, Eqn. \ref{Eqn:WeightUpdates} implies that the nodes closest to the BMU are dominantly aligned with the input feature, with the effect of the alignment attenuating on moving away from the BMU for a given time step. Neighborhood Distance Function is given by,  
\bea 
\phi (u, v, t) = \exp{\left(\frac{-[d(u,v)]^2}{2[\sigma(t)]^2}\right)},
\label{Eqn:NDF}
\eea
with $d(u,v)$ being the distance metric between the coordinates $u$ and $v$, and $\sigma(t)$ is the attenuating factor at time step t following exponential distribution that reads, 
\bea 
\sigma(t) = \sigma(0) \exp{(-t/\beta)},
\label{Eqn:Sigma}
\eea 
where $\sigma(0)$ is the initial attenuating factor and $\beta$ is the attenuation decay rate. 

We employ \textit{Batch Self-Organizing Map} (BSOM) to ensure efficient training over the entire dataset at once, potentially leading to faster model convergence and a more stable cluster boundary. The input features are scaled using the {\tt MinMaxScaler} normalization scheme so that every value for a given feature ranges between 0 and 1, and all input features receive equal importance in the model training. Unlike traditional SOMs that update weights after considering each data point, BSOMs update weights after processing a batch of data. Treating the entire training dataset as a single batch impedes the model convergence, and varying the batch-size we find that batches of 500 events yield faster convergences and stable outputs. Prioritizing the simplicity of the model, we choose linearly decreasing values for the learning rate ($\eta$) and the attenuating factor ($\sigma$) with each batch. Starting from 1, we varied $\eta$ to 0, and $\sigma$ from 20 to 1. We use the implementation of the SOM algorithm provided in the {\tt GitHub} repository \cite{som:github}, with the necessary modifications mentioned above.

%%%(Reference: https://quantdare.com/self-organizing-maps-for-clustering/)

%%%%%%%%%%%%%%%%%%%%%%%%%%%%%%%%%%%%%%%%%%%%%%%%%%
%\input{Evgen}
\section{Event Generation and Reconstructed Objects}  \label{Event_gen}

This section provides an overview of the action plan followed for the preparation of input data required for training the Self-Organizing Map architecture. We sketch below the steps undertaken for event generations, object reconstructions, and feature selections:

\begin{itemize}
\item {\bf Generation of Monte Carlo Samples:} \\  
We have considered two dominant SM processes: $(i)$ associated production of a top-quark with a $W$-boson, with $W$-boson undergoing leptonic decay and the top decaying to a bottom quark and a hadronically decaying $W$ boson, and $(ii)$ production of di-jet QCD events for training the network. In order to achieve the boosted regime, we simulate the processes putting a generation level cut of 350 GeV on the minimum transverse momentum of the top quark and QCD parton (quark/gluon). We use {\tt MadGraph} (version 3.4.1) \cite{Alwall:2011uj} to generate hard scattering events at the leading-order (LO) in $\alpha_s$ followed by the desired decays using MadSpin. The collision energy has been considered to be 13 TeV, and both the renormalization and factorization scales are fixed at $H_T$/2, where $H_T$ denotes the cumulative transverse energy of the colliding partons. We consider the Parton Distribution Function (PDF) given by {\tt NNPDF23-lo-as-0130-qed} \cite{NNPDF:2014otw}. The generated events are passed through {\tt PYTHIA8} (version 8.310) \cite{Bierlich:2022pfr} to account for the initial and final state radiations, parton showering and hadronization. The HepMC \cite{hepmc} output obtained from {\tt PYTHIA8} is further fed into {\tt Delphes} (version 3.5.0) \cite{delphes} to consider the detector effects, using the CMS card.  

For the signal events, we follow a model-independent, generic framework \cite{Buchkremer:2013bha} leading to the rare top decay $\left( t \to cH \right)$, and use {\tt FeynRules} \cite{feynrules1, feynrules2} to generate the UFO output, which is provided to MadGraph to generate hard scattering events. 

\item {\bf Object Reconstruction:} \\ 
We use {\tt FastJet} \cite{fastjet1, fastjet2} with anti-kT algorithm \cite{antikt} for the reconstruction of jets, with jet radius set at $R = 1.0$, and optimized based on the truth-level event information to capture the decay products of the top quark, aligning the jet-axis along the direction of the top. To mitigate contamination from underlying events (UE) and multiple parton interactions (MPI), jet trimming \cite{Krohn:2009th} is applied. This involves reclustering the enveloped products into subjets with jet radius 0.3 using kT algorithm, and removing the subjets with momentum fraction less than 3\% of the large-R jet momentum. Reconstructed large-R jets with $p_T >$ 400 GeV and mass $\in$ [140, 200] GeV are finally selected for further analysis; these are the candidate top-jets. 

For the identification of flavor-tagged subjets, we identify b- and c-hadrons within the large-R jet from particle-level information. A subjet is labeled b- or c-tagged if the angular separation $(\Delta R)$ between the corresponding hadron and the subjet-axis is less than the subjet radius. Detector effects are considered for the flavor-tagging by imposing b- and c-tagging efficiencies to be 70\% and 40\% respectively, and using a $p_T$-dependent formula for mistagging probabilities \cite{ATLAS:2017bcq,ATLAS:2015prs}. To prevent multiple tagging, b-tagging is performed first on $p_T$-ordered subjets, and the tagged subjets are excluded from subsequent flavor-tagging. Subjets failed to be flavor-tagged are labeled as light jets.

\item {\bf Selection of Jet Substructure based Observables as Input Features:} \\
We have followed the approach outlined in \cite{Chowdhury:2023jof} for the consideration of relevant jet substructure based observables, presenting a comprehensive list of the features used for SOM training in Table \ref{tab:inputs}. 

% ------------------------------------
\begin{table}[!htb]
\centering
\begin{tabular}{lllllll}   
\toprule
\multicolumn{7}{l}{\textbf{Input features for SOM}} \\  
\midrule
$p_{T}^{b-jet}$ & $\Delta E_T^{b-jet1}$ & \multirow{2}{*}{$\Delta E_T^{c-jet}$} & $\tau_{21}^{1}$ & $\Delta X_{cj}$ & $\Delta X_{bj}$ & $N_{c-jet}$ \\ [2mm]
$p_{T}^{c-jet}$ & $\Delta E_T^{b-jet2}$ &  & $\tau_{32}^{1}$ & $\Delta X_{cb}$ & $N_{b-jet}$ & $N_{Ljet}$  \\ 
%$N^{Displaced}_{Track}$ & $M_{DV}$ & $N_{DV}$ 
\bottomrule  
\end{tabular}
\caption{\label{tab:inputs}{The set of input variables used for training SOM.}}
\end{table}
%---------------------------------------
We briefly sketch an outline of few important subjet based features below, referring the interested readers to \cite{Chowdhury:2023jof} for further details: 
%-----
\begin{itemize} 
\item [-] {\underline{N-subjettiness}} ($\tau^{\beta}_{N}$): N-subjettiness characterizes the substructure of boosted jets \cite{NSubjettiness}. Considering N hypothetical subjet axes, N-subjettiness is determined by summing up the angular distances of jet constituents with respect to their respective closest subjet axes. Essentially, it indicates how well a jet aligns with N subjet-structure hypothesis. To establish the prominence of the existence of three subjet axes within the large-R jets of the signal (or the irreducible background) over two, and two subjet axes over one (for QCD di-jet background), we analyze the ratios of N-subjettiness measurements, specifically measuring $\tau_{N}^{\beta}/\tau_{N-1}^{\beta}$, labeled $\tau_{N (N-1)}^{\beta}$, for respectively $N = 3, ~2$, considering the angular weightage exponent $\beta = 1$. These ratios enables the isolation of the QCD multijets from top jet samples. 

\vspace*{0.25cm}
\item [-] {\underline{Subjet Multiplicities \& Transverse Momenta}}: Multiplicities of light and flavor-tagged subjets $\left( N_{b-jet}, ~N_{c-jet} \right)$ along with the transverse momentum of leading flavored subjet $\left( p_{T}^{b-jet}, ~p_{T}^{c-jet} \right)$ are crucial for the identification of signal and different SM backgrounds.

\vspace*{0.25cm}
\item [-] {\underline{Energy- and Mass-Fraction Information}}: 
{{To measure how the energy of a large-R jet is distributed among its subjets, we looked at the fraction of its total energy contained within its constituent subjets}}. For a given subjet, this is measured as, $\Delta E_{T}^{subjet} = \frac{E_{subjet}}{E_{topjet}}$.
We further estimate the composition of large-R jet, with an observable measuring the fractional mass retained after removing the invariant mass of specific subjet combinations (indexed by $i,~j$): $\Delta X_{ij}  = \left( 1 - \frac{M_{ij}}{M_{topjet}} \right) $.
Accurate identification of light jets is crucial for this analysis, and is achieved through a $\chi ^{2}$-analysis detailed in \cite{Chowdhury:2023jof}. 
\end{itemize}
%------
\end{itemize}

%%%%%%%%%%%%%%%%%%%%%%%%%%%%%%%%%%%%%%%%%%%%%%%%%%%%%%
\section{Results}      \label{Result}

Having trained an SOM with QCD di-jet and SM top decay $( t \to bW)$ samples, we discuss our observations in this section.

\subsection{Training on Data and the U-Matrix:} \label{train_2clust_som}

\begin{figure}[!htb] 
\centering   
{\includegraphics[width=0.4\textwidth]{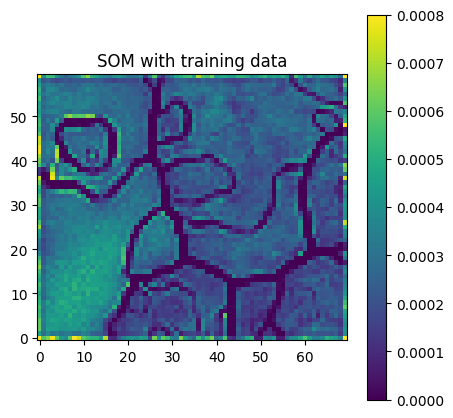}  }
{\includegraphics[width=0.38\textwidth]{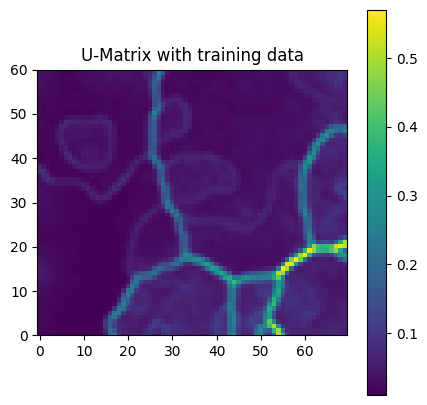}  }
\hfill
\caption {Left: The trained SOM, displaying clusters formed from top decay and QCD multijet events. Black lines indicate regions with no data points, separating distinct clusters. Right: The U-Matrix visualization of the trained SOM, revealing a vertical boundary dividing the map in a roughly 1:2 ratio, with the larger right-hand section showing further sub-clusters.}  
\label{Fig:SOM-Umat-2clust}
\end{figure}

A Self-Organizing Map (SOM) with a $60 \times 70$ grid is trained over 500,000 events, each event represented by thirteen features listed in Table \ref{tab:inputs}. The vector weights over each node are initialized at random. Taking into account each event, the comparison between the event feature vector and the vector weights is drawn on each node, the Best Matching Unit is identified, and the weight vectors are modified according to Equation \ref{Eqn:WeightUpdates}. Through multiple iterations on the training dataset, clusters appear and distinct cluster boundaries emerge, as shown in the left panel of Figure \ref{Fig:SOM-Umat-2clust}. The cluster boundaries, which are the darker regions of the grid, represent areas without data points. They effectively envelope the brighter regions, which capture events with similar underlying traits. Moving forward, each cluster would unfold a unique predominant event topology encountered in the training dataset. The intensity of the color reflects the density of the events on the map, scaled by the total number of events used to train the grid.

We determine the strength of the identified cluster boundaries with a U-Matrix. The elements of U-Matrix are calculated for every node by measuring the average Euclidean distance between the weight vector on the concerned node and that on its neighboring nodes. This visualization allows us to identify the most significant cluster separations. The U-Matrix that emerges from our trained SOM is presented in the right panel of Figure \ref{Fig:SOM-Umat-2clust}. The U-Matrix separates the grid with a vertical boundary leading to 2 major clusters, along with 5 smaller, more tightly defined sub-clusters, within the cluster at right. The detailed analysis of these boundaries and the event populations within each cluster is presented in the following sections.

%%%%%%%%%%%%%%%%%%%%%%%%%%%%%%%%%%%%%%%%%%%%%%%%%%%%
\subsection{Robustness of the cluster boundaries}
\begin{figure}[!htb] 
\centering
{\includegraphics[width=0.5\textwidth]{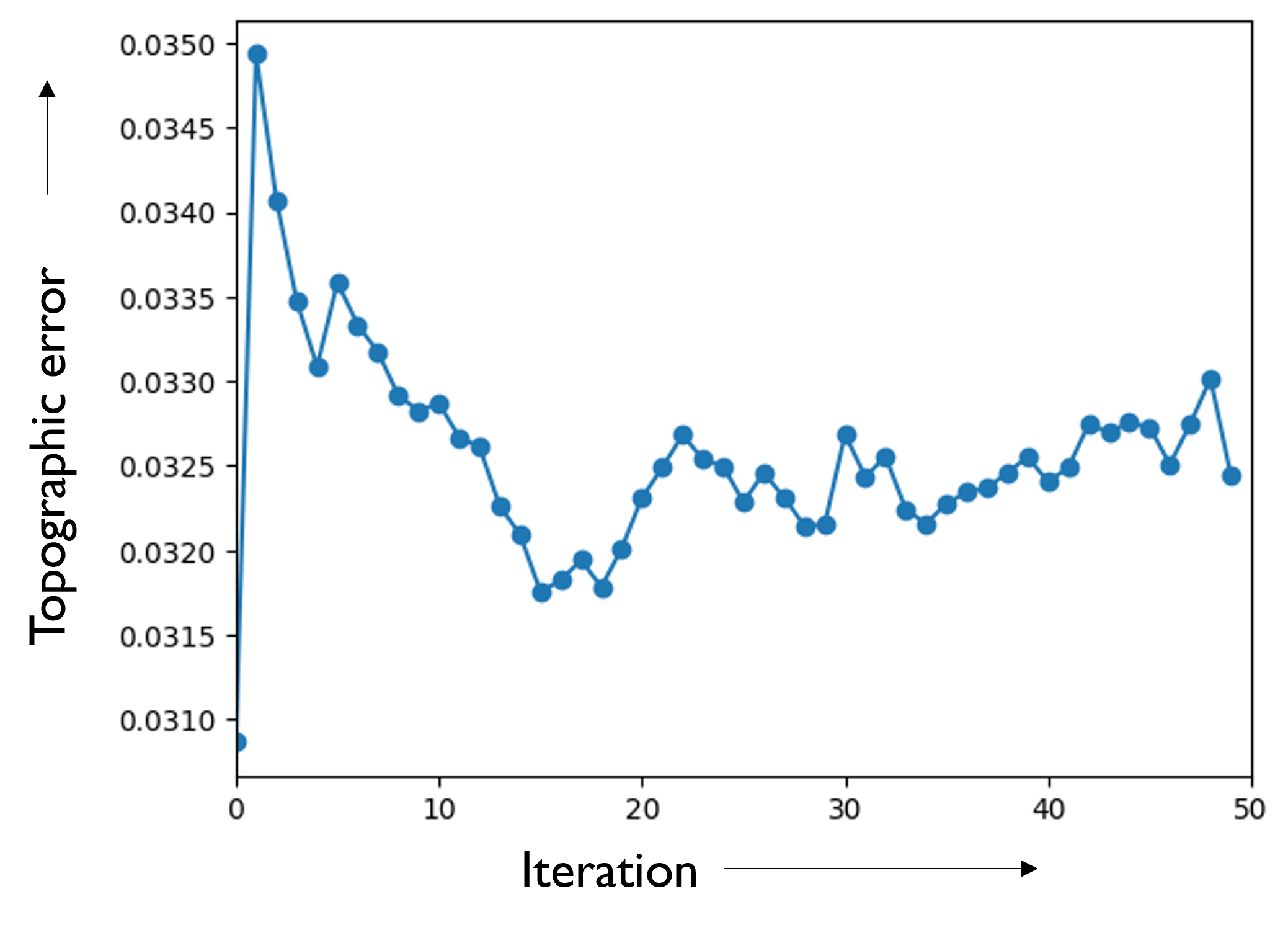}  }
\hfill
\caption { Topographic error after training the map with QCD multijet and SM top decay sample for 50 iterations. }  
\label{Fig:topoerr_2clust}
\end{figure}

We have taken a step further to assess the topographic error for the obtained map, which determines the fraction of events for which the Best Matching Unit and second Best Matching Unit are adjacent to each other. Figure \ref{Fig:topoerr_2clust} illustrates the evolution of the topographic error in the first 50 iterations. The error starts at 3.5\% and stabilizes around 3.3\%, indicating an effective model convergence.
%
%---------------------------------------
%---------------------------------------
%
To further assess the quality of the clustering, we employ two evaluation metrics: the Silhouette score \cite{ROUSSEEUW198753} and the Davies-Bouldin Index (DBI) \cite{DBI}.
\begin{itemize} 
\item {\underline{Silhouette score:}} The Silhouette score measures how well an event fits within its assigned cluster compared to other clusters. For a given event \textit{e}, is defined as:
\[
S = \frac{b - a}{\max(b, a)}
\]
where, `$a$' denotes the mean distance between \textit{e} and all other events within its own cluster, and `$b$' is the mean distance between \textit{e} and all events in its nearest-neighboring cluster. The average value of $S$ for all events gives the average Silhouette score. This score ranges between $[-1, 1]$, where positive values indicate moderate to good clustering and negative values suggest poor clustering. A score close to zero implies that the events are globally populated in the vicinity of the cluster boundaries. 

\item {\underline{Davies-Bouldin Index (DBI)}}: This metric is used to assess the quality of the clustering. A lower DBI value signifies the presence of compact, well-separated clusters. DBI is given by,
\[
\text{DBI} = \frac{1}{N} \sum_{i=1}^{N} \max_{i \neq j} \left( \frac{S_i + S_j}{D_{ij}} \right)
\]
where, $N$ is the total number of clusters, $S_i$ is the mean intra-cluster distance for the $i^{th}$ cluster (a measure of compactness), and $D_{ij}$ denotes the distance between the centroids of the clusters $i$ and $j$ (a measure of separation). The condition $\max_{i \neq j}$ allows the identification of the least compact cluster $j$ and close to the cluster $i$ (worst case similarity).
\end{itemize} 

We implement these two metrics from \texttt{sklearn} library \cite{scikit-learn} to evaluate the clustering performance achieved after training. We obtain a Silhouette score of 0.46, which indicates that the trained model achieves moderate clustering quality, with reasonably well-defined clusters. A DBI score of 1.04 supports this claim, suggesting fair cluster compactness and separation. These metrics confirm satisfactory model performance on the training dataset. For cross-verification, we compared these metrics evaluated for the trained SOM with those obtained from the k-means clustering. The result is in close agreement and demonstrates
consistency and robustness.

%----------------------------------------------------

\subsection{Analysis of the Weight Planes and Clusters}

\begin{figure}[!htb]
    \centering
    \includegraphics[width=0.5\linewidth]{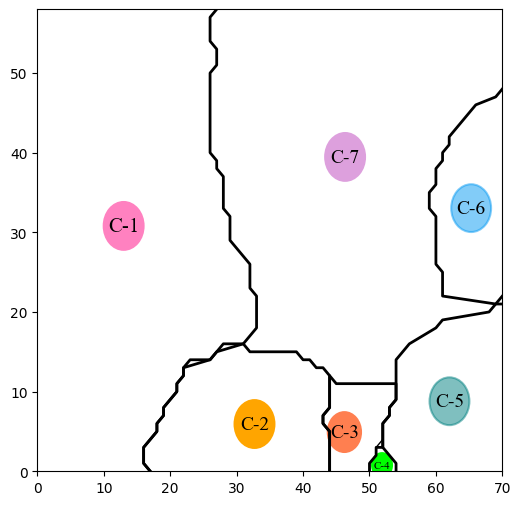}
    \caption{The prominent clusters as seen from Figure \ref{Fig:SOM-Umat-2clust}. The clusters are assigned with specific numbers for convenience. The events inside each of these clusters are then identified for further analysis.}
    \label{fig:cluster}
\end{figure}
%-----------------------------------
As stated in Section \ref{algo}, SOM provides a unique data visualization tool to analyze the underlying structure of the training data, without any prior knowledge of its topography. Basically, SOM clusters the data points (events) by identifying the underlying structure and grouping similar data points together. Therefore, the {\it cluster boundary}, functionally, resembles a gradient where the neighboring nodes of the grid represent progressively different data traits. Based on the representation of the U-matrix in the right panel of Figure \ref{Fig:SOM-Umat-2clust}, we identify seven (7) clusters leveled as C-1 to C-7, as shown in Figure \ref{fig:cluster}, to proceed with further analysis.  

An insight into the weight planes would provide a better understanding of the clusters and their boundaries. The ``weight'', in this context, carries an entirely different meaning than its conventional sense for Artificial Neural Networks (ANN). For ANNs, the values corresponding to each input feature are multiplied by a weight, and their total contribution is measured by passing them as an argument for an activation function. For SOMs, the weights are assigned to the output nodes, without any activation function. Through adjustments of these weights, SOM creates insights into the feature space. Hence, to gain a deeper understanding of how these features shape different collision outcomes, we now discuss their corresponding weight planes. 

\begin{figure}[!htb]
\includegraphics[width=0.3\textwidth]{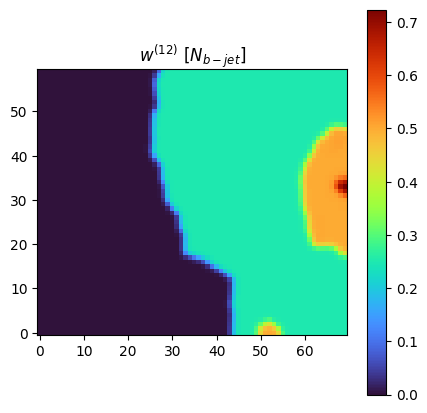} 
\includegraphics[width=0.3\textwidth]{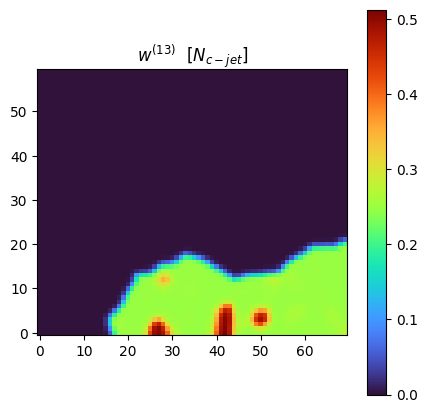}  
\includegraphics[width=0.3\textwidth]{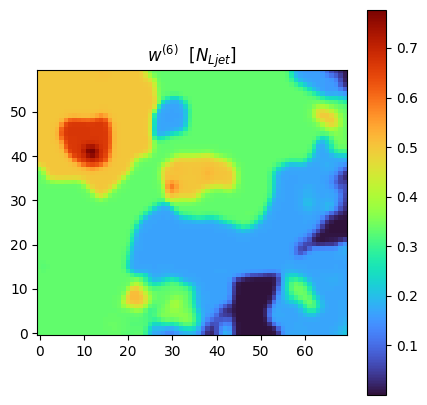} 
\includegraphics[width=0.3\textwidth]{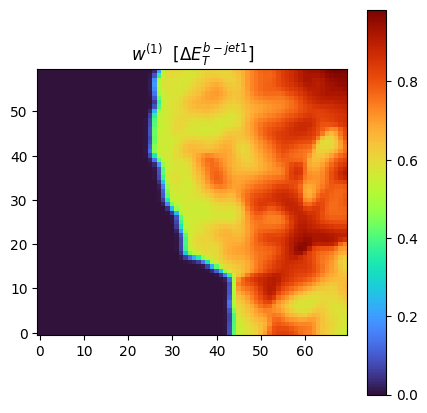} \hfill
\includegraphics[width=0.3\textwidth]{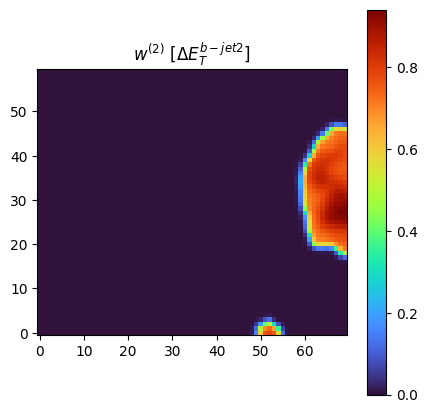} \hfill 
\includegraphics[width=0.3\textwidth]{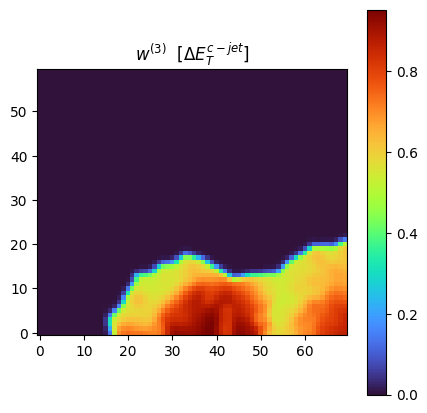}  
%{\includegraphics[width=0.3\textwidth]{fig/W9_pTb.png}  }
%{\includegraphics[width=0.3\textwidth]{fig/W10_pTc.png}  }
\caption{Weight vector plane corresponding to the features, $N_{b-jet}$, $N_{c-jet}$, $N_{Ljet}$, $\Delta E_T^{b-jet1}$, $\Delta E_T^{b-jet2}$, $\Delta E_T^{c-jet}$, respectively. The unified effect of these weights is what we see in the U-Matrix in Figure \ref{Fig:SOM-Umat-2clust}.}  \label{Fig:W-map1}
\end{figure}
%-------------------------------------------------

In the upper panel of Figure \ref{Fig:W-map1}, we display the weight planes corresponding to different jet multiplicities, namely, the number of b and c-tagged jets and light jets. The clearly visible boundaries in the weight planes for $N_{b-jet}$ and $N_{c-jet}$ highlight the role these two features play in the clustering process. For example, the dark region on the left side of the weight plane of $N_{b-jet}$ represents the grids that correspond to events having negligible number of b-tagged jets - a well-known feature of the QCD di-jet events. The map of the weights of $N_{c-jet}$ supports this claim, as these events are also expected to have a very small number of c-tagged jets. Events with non-zero values of both b- and c-tagged jet multiplicities are mapped to the lower-right corner of the map. As expected, light jet multiplicity did not play much of a role as both the QCD and top events contain a large number of light jets. Moving to the lower panel of Figure \ref{Fig:W-map1}, where we show the map of weight planes for three other decisive features, namely $\Delta E_T^{b-jet1}$, $\Delta E_T^{b-jet2}$, and $\Delta E_T^{c-jet}$. As seen in the figures, these features help to understand the energy profile among the flavor-tagged subjets. These observables are also mapped with visible decision boundaries in the weight planes, significantly contributing to the different clusters discussed in Figure \ref{fig:cluster}. 

%-------------------------------------------------
\begin{figure}[!htb] 
\centering   

{\includegraphics[width=0.3\textwidth]{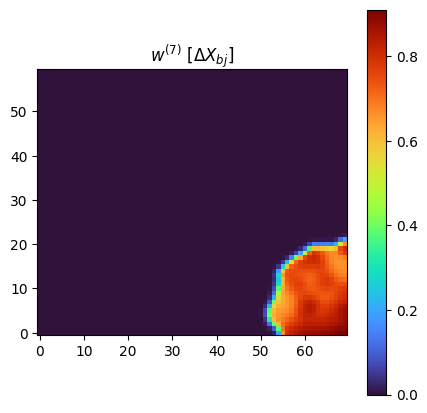}  }
{\includegraphics[width=0.3\textwidth]{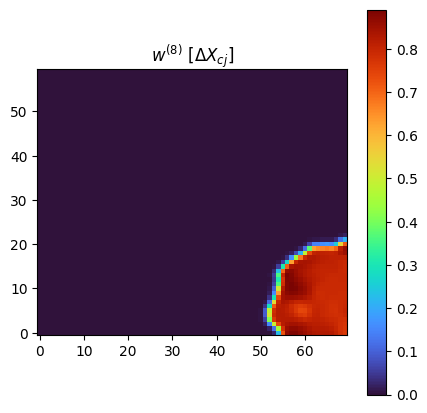}  }
{\includegraphics[width=0.3\textwidth]{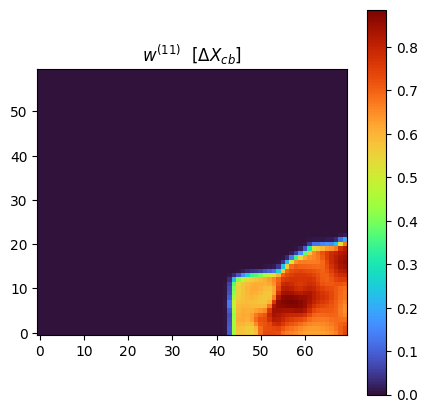}  }
\caption{Weight plane corresponding to the features, $\Delta X_{bj}$ , $\Delta X_{cj}$, and $\Delta X_{cb}$ respectively.}
\label{Fig:W-map2}
\end{figure}
%-------------------------------------------------

The presence of a W boson within the large R top jet motivated us to include some observables in terms of the invariant mass of the flavor-tagged subjets, namely $\Delta X_{bj}$, $\Delta X_{cj}$, and $\Delta X_{cb}$. As evident in Figure \ref{Fig:W-map2}, along with the multiplicities of the jet and their energy profile, these features greatly help clustering events. In fact, the brightest boundaries in the U-Matrix are primarily determined using the multiplicities of light-jet, b-jet, and c-jet, along with the energy and mass distributions among these subjets. 

The significant observations obtained through the weight plane analysis are as follows: 
\begin{itemize} 
\item The leftmost cluster (\textit{viz.}, C-1) includes mostly the events having high light-jet multiplicity with (almost) no b- or c-tagged jets. Therefore, this cluster certainly corresponds to the QCD multijet events. 

\item The right subclusters (other than C-1) have events with one or more b-tagged jets and light-jets. Additionally, the events mapped at the bottom right corner (\textit{viz.}, C-2 to C-5), in addition, possess non-zero c-tagged jets. Therefore, the right-hand cluster corresponds to the events from the top decay. Clusters with 2 or more b-tagged jets would include events originating from non-standard decay of the top quark.  

\item  There exist at least two clusters, namely C-4 and C-6, which
contain events that have two or more b-tagged jets along with at least one c-tagged jet. These events are not expected to originate from the SM decay of the top quarks and, therefore, could be considered as clusters having ``anomalous'' events.  

\end{itemize}

%%%%%%%%%%%%%%%%%%%%%%%%%%%%%%%%%%%%%%%%%%%%%%%%%%%%%%%%%%%%%
\subsection{Clustering efficiency with Test dataset}

To assess the performance of the SOM training, we test the model with three kinds of samples,
\begin{enumerate}
    \item {\underline{Test dataset1}}: $p p \to tW$ events with the top following the dominant decay mode producing b- and c-quarks in the final state ($t \to b W^+$ with $W^+ \to c\bar{s}$ and the corresponding charge-conjugated process). The $W$-boson produced in association with the top quark decays leptonically. 
    
    \item {\underline{Test dataset2A/2B/2C}}: $p p \to tW$ events with top quark decaying through an FCNC decay, $t \to c H$ followed by $H \to b\bar{b}$/$H \to W W^*$/$H \to ZZ^*$ decay. The $W$-boson produced in association with the top quark decays leptonically. 
    
    \item {\underline{Test dataset3A/3B}}: The most dominant channel, $p p \to t\bar{t}$, with one of the tops decaying leptonically\footnote{ The simulation methodology for the $t\bar{t}$ events, including the software suites, is identical to that described in Section \ref{Event_gen}.} $(t \to bW, W \to \ell \nu_{\ell})$, and the other top undergoes $t \to bW$ decay or $t \to cH$ decay with $W$/$H$ decaying hadronically.
\end{enumerate}
A summary of the test data sets is shown in Table \ref{tab:Test_Dataset}. 

%-------------------------------
\begin{table}[!htb]
\centering
    \begin{tabular}{c|c} \hline
    Dataset & Process considered \\ \hline
Test dataset1  & $pp \to Wt ~(p_T^{t} >350 ~GeV),\;  W \to \mu ~\nu_{\mu}; $ \\ 
     (SM)  & $ t \to bW,\; W \to c\;s$ \\ \hline 
Test dataset2A  &  $pp \to Wt ~(p_T^{t} >350 ~GeV),\;  W \to \mu ~\nu_{\mu};$ \\ 
      (BSM)  &  $t \to cH,\; H \to b\;\bar{b}$ \\ \hline
Test dataset2B  &  $pp \to Wt ~(p_T^{t} >500 ~GeV),\;  W \to \mu ~\nu_{\mu}; $\\ 
      (BSM)  &  $t \to cH,\; H \to W W^{*},\; W \to j j$ \\ \hline
Test dataset2C  &  $pp \to Wt ~(p_T^{t} >500 ~GeV),\;  W \to \mu ~\nu_{\mu};$\\ 
      (BSM)  &  $t \to cH,\; H \to Z Z^{*},\; Z \to j j$ \\ \hline
    \hline
$\rm Test dataset3A $ & $pp \to t\bar{t} ~(p_T^{t} >350 ~GeV),\; t \to bW,\;  W \to \mu ~\nu_{\mu}; $ \\ 
     (SM)  & $ \bar{t} \to bW,\; W \to j\;j $ \\ \hline
$\rm Test dataset3B $ &  $pp \to t\bar{t} ~(p_T^{t} >350 ~GeV),\;  t \to bW,\;  W \to \mu ~\nu_{\mu};$ \\ 
      (BSM)  &  $\bar{t} \to cH,\; H \to b\;\bar{b}$ \\ \hline
    \end{tabular}
\caption{The details of the three types of dataset considered in testing the trained SOM.}    \label{tab:Test_Dataset}
\end{table}
%--------------------------------

\begin{figure}[!htb]
\centering
\includegraphics[width=0.4\textwidth]{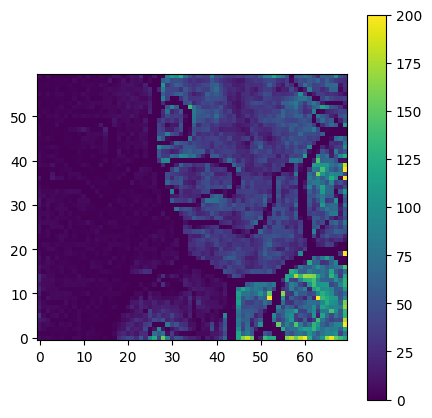} 
\includegraphics[width=0.4\textwidth]{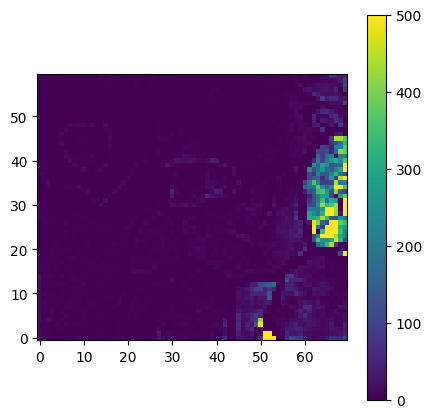} 
\caption{ Left: The projection of the events corresponding to the two test datasets - "Test dataset1" (left panel) and "Test dataset2A" (right panel). We refer the main text for more details. }
\label{fig:test_2clust}
\end{figure}

For validation, we calculate the efficiencies for independent (unseen) datasets of the SM top and QCD di-jet events. Almost 90\% of the QCD di-jet samples accumulated in the left clusters (e.g., C-1 and C-2) of the map, while 90\% of the SM top samples are assigned to the right clusters (C-3 to C-7) as portrayed in the first two rows of Table \ref{tab:cluster}. We next proceed with the analysis using the test datasets. Figure \ref{fig:test_2clust} shows the clustering of the two test datasets, namely Test dataset1 and Test dataset2A, in respectively left and right figures. The weight plane analysis suggests that Test dataset1 is dominantly clustered in the right half of the SOM grid, as these regions correspond to non-zero multiplicities of both b- and c-tagged jets. On the other hand, for Test dataset2A, which has higher multiplicities of b-tagged jet, should be clustered primarily in C-4 and C-6. Both figures confirm our expectation, and therefore motivate us to calculate the cluster efficiencies, defined as the number of events captured in a given cluster over the total number of events analyzed, for each cluster for a given test dataset. We select 100,000 events for each test dataset that satisfy the pre-selection cuts and trigger requirements, and tabulate the efficiencies in Table \ref{tab:cluster}. The higher efficiencies in C-5, C-6 and C-7 for Test dataset1, and C-4 and C-6 for Test dataset2A, clearly reflect the performance of the training algorithm. The phase space of the other two test datasets corresponding to BSM physics, namely Test dataset2B and Test dataset2C, are expected to have strong overlap with the QCD multijet events. As the numbers in the last two rows of Table \ref{tab:cluster} reveal, these events are indeed clustered in C-1 and C-2, as expected. Furthermore, we observe that $t\bar{t}$ events, like single top quark events, follow expected patterns. Test dataset3A characterized by the dominant top decay mode, are principally captured in cluster C-7. In contrast, events for Test dataset3B are distributed among the clusters C-4, C-6, and C-7, revealing a more diverse event topology, as shown in the last two rows of Table \ref{tab:cluster}. Based on these observations, it is reasonable to expect that a combined analysis of $tW$ and $t\bar{t}$ production processes would improve the achievable limit on the top FCNC decay.

%------------------------------
%-----------------------------

%-----------------------------
\begin{table}[!htb]
    \centering
  % \hspace{-2cm}
    \begin{tabular}{|c|c|c|c|c|c|c|c|}\hline
     Process  & C-1 & C-2 & C-3 & C-4 & C-5 & C-6 & C-7 \\ \hline 
     \hline
    SM top   & 5.45\% & 4.83\% & 2.51\% & 0.21\% & 12.19\% & 6.81\% & {\bf 67.15\%}\\ \hline
    QCD di-jet& {\bf 82.84\%} & 7.69\% & 0.23\% & 0.03\% & 0.47\%&1.91\% &6.26\%\\ \hline \hline 
   Test dataset1& 4.48\% & 5.39\% & 5.04\% & 0.42\% &{\bf 22.48\%} &11.02\%& {\bf 49.62\%} \\ \hline
    Test dataset2A & 2.19\% & 0.88\% & 3.65\% & {\bf 12.45\%} & 7.67\%& {\bf 61.06\%} &7.99\%\\ \hline
     Test dataset2B  & {\bf 46.93\%} & {\bf 36.78\%} & 0.36\% & 0.035\% & 2.67\% & 0.51\% & 11.08\%\\ \hline
     Test dataset2C & {\bf 49.06\%} & {\bf 35.37\%} & 0.45\% & 0.049\% & 2.22\% & 0.67\% &10.96\%\\ \hline 
     \hline
      Test dataset3A &4.1\%  & 4.86\% & 2.44\% &0.1\% &11.5\% & 5.5\% & {\bf 70.77\%} \\ \hline
      Test dataset3B &0.9\% & 1\% &3.5\%  & {\bf 12.22\%} &7.9\% & {\bf 56.42\%} & {\bf 15.27\%} \\ \hline
    \end{tabular}
    \caption{Cluster efficiencies, defined as the number of events clustered in the given cluster over the total number of events analyzed, corresponding to each cluster for a given test dataset. The dominant contribution(s) of each dataset towards the clusters are expressed in bold letters.}     \label{tab:cluster}
\end{table}
%--------------------------

%%%%%%%%%%%%%%%%%%%%%%%%%%%%%%%%%%%%%%%%%%%%%%%%%%%%%%%%%%%%%%%%%%%%
\section{BSM Reach: Event Selection to Signal Significances}   \label{BSM}

We now proceed to estimate the reach of the BSM physics process, viz. the FCNC decay of the top quark via $t \to c H$ process, at the high-luminosity run of LHC. To identify the new-physics rich clusters (or, equivalently, the clusters with anomalous events), we first apply a pre-selection cut (trigger requirement): we select events with at least one lepton (electron/muon) with $p_{T} > 250$~GeV and missing transverse energy $\cancel{P_T} >$ 200 GeV. This choice will ensure that we are indeed focusing on the phase space when the top quarks are produced with sufficient boost.

 %---- higgs reconstruction 
\begin{figure}[!htb]
  %  \centering
    \includegraphics[width=0.4\linewidth]{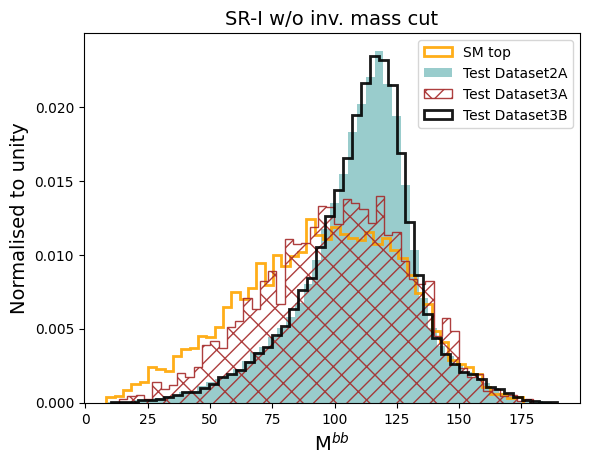}
    \hspace{1cm}
     \includegraphics[width=0.4\linewidth]{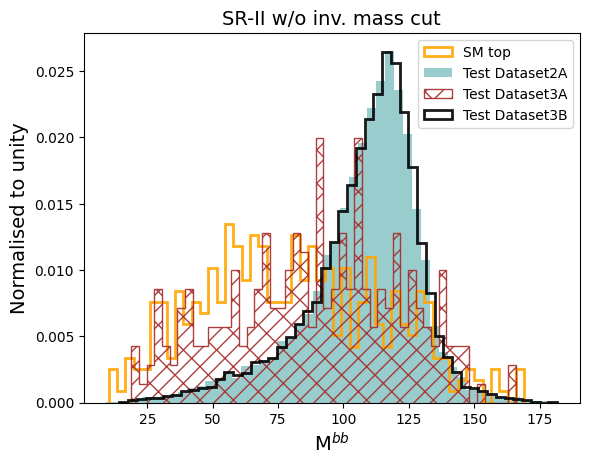}
     
    \caption{The invariant mass distribution of the two b-tagged jets for SR-I (left) and SR-II (right) without the invariant mass cut. }
    \label{fig:higgsreco}
\end{figure}
%------------------------------------

The first signal region (SR-I) is almost the same as ATLAS {and CMS study} \cite{ATLAS:2018jqi,CMS:2021gfa}, with the only difference being that the requirements are made at the subjet level. We extend the selection strategy of SR-I to SR-II, accommodating the further requirement of c-tagged jets within the large-R jet. It is needless to say that SR-II is expected to be the most efficient selection strategy in probing the top FCNC decay we are interested in. Finally, we also study the sensitivity of a relatively relaxed criterion to the multiplicity of b-tagged jets within the large R jet by introducing the third signal region, SR-III. Note that, for both SR-I and SR-II, the presence of at least two b-tagged jets enables us to calculate the invariant mass of the two b-tagged jets, and compare the reconstructed mass with the Higgs boson mass. In Figure \ref{fig:higgsreco}, we show the distribution of the invariant mass mentioned above for the SM Top samples, Test dataset2A, Test dataset3A and Test dataset3B that satisfies the criteria for SR-I and SR-II without the invariant mass cut. For SR-I and SR-II, therefore, we further demand that the invariant mass of the two leading b-tagged jets satisfy the following: $M_{bb} = [110,140]~{\rm GeV}$. The details of the signal regions are listed in Table \ref{tab:signal region}. 

%-------------------------------
%%% Signal region 
\begin{table}[!htb]
\centering
\begin{tabular}{|c|c|}\hline 
SR-I & nSubjet $\ge$ 3, ~ $N_{b-jet} \ge$ 2, ~ $M_{bb} = [110,140]~{\rm GeV}$\\ \hline
SR-II & nSubjet $\ge$ 3, ~ $N_{b-jet} \ge$ 2,  ~ $N_{c-jet} >$ 0, ~ $M_{bb} = [110,140]~{\rm GeV}$ \\ \hline
SR-III & nSubjet $\ge$ 3, ~ $N_{b-jet} >$ 0, ~$N_{c-jet}>$ 0 \\ \hline
%SR-IIIb & nSubjet $\ge$ 3, ~ nb $>$ 0, ~nc = 0 \\ \hline
    \end{tabular}
\caption{Signal regions defined to probe the top FCNC decay. Here the observables $nSubjet$, $N_{b-jet}$ and $N_{c-jet}$ denote the total number of subjets, number of b-tagged jets and number of c-tagged jets within the large-R jet, respectively. The invariant mass of the two leading b-tagged jets is denoted by the quantity $M_{bb}$. }
\label{tab:signal region}
\end{table}
%-------------------------------

As noted previously, the FCNC interactions between the top quark and the Higgs boson via the $t-c-H$ coupling are essential in our quest for BSM physics. To evaluate the feasibility of detecting this exotic mode in the boosted regime using the SOM algorithm, we compute the signal significance, defined as 
$\mathcal{S}  = \sqrt{2 \left[ (S + B) \ln\left(1 + \frac{S}{B}\right) - S \right]}$  \cite{Cowan:2010js} 
where S and B represent the number of signal and background events that meet the jet selection criteria and satisfy the trigger/pre-selection requirement. Before we proceed to calculate the signal significances for the three signal regions, it is important to estimate the cluster efficiencies for each of the SRs. In Table \ref{tab:xsec}, we show the leading order (in $\alpha_s$) production cross section for all the processes considered here with $\sqrt{s} = 13 ~{\rm TeV}$. The last column in Table \ref{tab:xsec} estimates the expected number of events at the 13 TeV run of LHC with integrated luminosity $\mathcal{L} =3000~fb^{-1}$. It is important to note that even though QCD di-jet events seem to overshadow signal events as seen in $3^{rd}$ column of $2^{nd}$ row, the number of events that eventually satisfy the trigger and jet selection cuts, and contribute to the signal regions defined in Table \ref{tab:signal region} is negligible.  

%----------------------------------
\begin{table}[!htb]
\centering
    \begin{tabular}{|c|c|c|c|}
    \hline
      Process & Cross-section ($fb$) & $\mathcal{BR}$ & Cross-section$\times \mathcal{BR} \times \mathcal{L}$ \\ \hline
       SM top & 435 & $0.11\times1.0\times0.68$ & 9.76$\times10^{4}$ \\ \hline
       QCD di-jet & $3.91 \times 10^{6}$& 1  & 1.17$\times10^{10}$ \\ \hline
        Test dataset2A & 435 & $0.11\times1.0\times0.6$ & 8.61$\times10^{4}$\\ \hline
        Test dataset2B & 72.3 & $0.11\times1.0\times0.215\times0.674^2$ & 2330\\ \hline
        Test dataset2C & 72.3 & $0.11\times1.0\times 0.0264\times0.547^2$ & 188.5\\ \hline
        Test dataset3A & 7879 & $0.11\times1.0\times0.68\times1.0$ &$1.77\times10^{6}$\\ \hline
        Test dataset3B & 7879 &  $0.11\times1.0\times0.6\times1.0$ &$1.56\times10^{6}$\\ \hline
    \end{tabular}
    \caption{ The production cross section and branching ratio for the different processes considered in this analysis. The last column denotes the expected number of events at the 13 TeV run of LHC with integrated luminosity $\mathcal{L} =3000~fb^{-1}$.}
    \label{tab:xsec}
\end{table}
%----------------------------------

In order to calculate the signal significance, we estimate the number of events captured in each cluster region for the BSM test datasets that pass the trigger and jet selection cuts. In Tables \ref{tab:SR1} - \ref{tab:SR3}, we show the effective number of events that are clustered in each of those clusters that satisfy all the selection cuts for SR-I to SR-III. The notation S2A, S2B, S2C and S3B denotes the effective number of events for the four test datasets, \textit{viz.}, Test datasets 2A, 2B, 2C and 3B, respectively. 

%----------------------------------
\begin{table}[!htb]
   \centering
    \begin{tabular}{|c|c|c|c|c|c|c|c|}\hline
   & C-1  & C-2 & C-3 & C-4 & C-5 & C-6 & C-7 \\ \hline 
   B1 & - & - & - & 55.6 & 8.8 &1673.8 & - \\ \hline 
    B2 & - & - & - & 796.5 &194.7&25346.4 & - \\ \hline 
    B & - & - & - & 852.1 &203.5 & 27020.2 & - \\ \hline 
    \hline
    S2A & - & - & 0.9  & 6426.5 & 539.8 & 16229.9 &- \\ \hline 
  S2B & -  & - & -  &0.33 & - &  -  &-  \\
 \hline 
  S2C & -  & - & -  &0.025 & -  & - & -\\
 \hline
 S3B & - & - & 78 & 111196.8 &9141.6 & 282375.6 &- \\ \hline  
  S & - & -  & 78.9 &117623.7& 9681.4 & 298605.5 & - \\ \hline 
    \end{tabular}
\caption{ The effective number of events clustered in each cluster that satisfy all the selection cuts for SR-I. The notation `-' indicates that no events exist corresponding to the given cluster. }
    \label{tab:SR1}
\end{table}
%----------------------------------

%----------------------------------
\begin{table}[!htb]
   \centering
    \begin{tabular}{|c|c|c|c|c|c|c|c|}\hline
       & C-1  & C-2 & C-3 & C-4 & C-5 & C-6 & C-7 \\ \hline 
  B1 & - & - & - & 55.6 & 8.8 &- & - \\ \hline 
    B2 & - & - & - & 796.5 &194.7&- & - \\ \hline 
    B & - & - & - & 852.1 &203.5 & - & - \\ \hline 
    \hline
     S2A & - & - & 0.9  & 6426.5 & 539.8 & - &- \\ \hline 
  S2B & -  & - & -  &0.33 & - &  -  &-  \\
 \hline 
  S2C & -  & - & -  &0.025 & -  & - & -\\
 \hline
 S3B & - & - & 78 & 111196.8 &9141.6 & - &- \\ \hline  
  S & - & -  & 78.9 &117623.7& 9681.4 & - & - \\ \hline 
    \end{tabular}
    \caption{ The effective number of events clustered in each cluster that satisfy all the selection cuts for SR-II.}
    \label{tab:SR2}
\end{table}
%----------------------------------

%-----------------------------------
\begin{table}[!htb]
   \centering
    \begin{tabular}{|c|c|c|c|c|c|c|c|}\hline
   & C-1  & C-2 & C-3 & C-4 & C-5 & C-6 & C-7 \\ \hline 
 B1& - & - & 237.2 & 218.6 & 11795 & - &-\\ \hline 
 \hline
 B2& - & - &3062.1 & 2637.3 &203992.5 & - & - \\ \hline 
 B & - & - &3299.3 & 2855.9 &215787.5 & - & - \\ \hline 
    \hline
    S2A& - & - & 1195.1 & 10840& 6750.2 & -& -\\ \hline 
  S2B& - & - & 1.7 & 0.9 & 62 &- & - \\
 \hline 
  S2C& - & - &0.3 &  0.1 & 4.2 &- & -\\
 \hline 
  S3B & - & - &19968 & 181755.6 & 124066.8 & - &- \\ \hline 
 S& - & - & 21165.1 &192596.6 & 130883.2 &- &- \\ \hline 
    \end{tabular}
    \caption{ The effective number of events clustered in each cluster that satisfy all the selection cuts for SR-III.}
    \label{tab:SR3}
\end{table}
%----------------------------------------

Considering an example, we define S2A as, 
\begin{equation} 
S2A = \sigma \times BR \times \mathcal{L} \times \epsilon_{cl}, 
\end{equation}
where $\sigma$ denotes the cross section of the process, whereas combined branching fraction leading to the final states denoted as ${\rm BR}$ with $\mathcal{L}$ being the integrated luminosity. The quantity $\epsilon_{cl}$ denotes the number of events passing the pre-selection and trigger requirements for signal region for each cluster over the total number of test samples. The total effective number of events for the signal and backgrounds is labeled as S and B, respectively, with ${\rm S = S2A + S2B + S2C + S3B}$, and ${\rm B = B1 + B2}$ where B1 and B2 are the effective number of events corresponding to the two background datasets, i.e., SM top and Test dataset3A. 

%------------------------------------
\begin{table}[!htb]
\centering
\begin{tabular}{|c|c|c|c|c|}   \hline
Signal Region  & S & B & $\mathcal{BR}( t \to cH)$  &\(\mathcal{S} \,\) \\ \hline
 &  & & 0.1\% & 2.5  \\
SR-I & 425989.5 & 28075.8 & 0.05\% & 1.3  \\ 
& & &0.03\% &0.8 \\
\cline{1-5}
&  &  & 0.1\% &3.8\\
SR-II  & 127384 & 1055.6 & 0.05\% & 1.9 \\ 
& & & 0.03\%& 1.2 \\
\cline{1-5}
 &  &  & 0.1\% & 0.7\\
SR-III & 344644.9 & 221942.7 & 0.05\% &0.4\\ 
& & & 0.03\%&  0.2\\ \hline
\end{tabular}
\caption{ Signal significances for three SRs defined based on different choices of flavor tagging at the subjet level.}  \label{tab:signi}
%at 3000 ${\rm fb}^{-1}$ integrated luminosity.
\end{table}

We next proceed to calculate the signal significance using the total number of signal and background events with 3000 ${\rm fb}^{-1}$ integrated luminosity. Our findings are tabulated in Table \ref{tab:signi}. We calculate the signal significance for each signal region with different choices of $\mathcal{BR}( t \to cH)$. As evident from the table, SR-II provides the best sensitivity in probing the new physics contributions to the top FCNC decay rate. {One can observe that with 3000 ${\rm fb}^{-1}$ of collision data and using the SOM clustering algorithm to identify the suitable signal region (i.e., SR-II), the model achieves a signal significance of $\sim$ 2$\sigma$ for $\mathcal{BR}( t \to cH)$ = 0.05\%, which is competitive enough with the existing limits of the ATLAS and CMS collaborations at the 13 TeV run of LHC \cite{ATLAS:2018jqi,CMS:2021gfa}.} %\textcolor{red}{One can observe that with 3000 ${\rm fb}^{-1}$ of collision data, the SOM clustering algorithm provides a signal significance of $\sim$ 2$\sigma$ for $\mathcal{BR}( t \to cH)$ = 0.05\%, which is competitive enough with the existing limits of the ATLAS and CMS collaborations at the 13 TeV run of LHC. }

%====================================
\section{Summary and Outlook}  \label{summary}
Probes of processes involving rare decays driven by Flavor-Changing Neutral Current (FCNC) offer a powerful avenue for exploring Physics beyond the Standard Model. The rare decay, $t \to cH$, considered in this article, offers a sensitive probe of potential New Physics interactions, in particular, the $t-c-H$ coupling. In contrast to the supervised learning methodology presented in \cite{Chowdhury:2023jof}, we present a complementary, model-independent analysis based on the unsupervised learning technique, targeting the boosted regime of this decay. With judicious choices of jet-substructure based observables along with efficient flavor tagging of subjets, similar to the approach followed in \cite{Chowdhury:2023jof}, we achieve an enhanced signal significance over our previous analysis \cite{Chowdhury:2023jof}, which outperformed the performances of advanced ML-driven approaches, \textit{viz.}, GNN-based LorentzNet and Transformer-based classifiers.

In \cite{Chowdhury:2023jof}, we argued that the judicious choice of jet-substructure based observables provided an edge over the GNN-based models which train on the 4-vectors of the large-R jet constituents. With the jet substructures of the signal $\left( t \to cH, ~H \to b\bar{b} \right)$, bearing close resemblance to those of the irreducible background $\left( t \to bW, ~W \to c\bar{s} \right)$, this schema achieved relatively lower significance. 
The transformer-based approaches implement an attention mechanism, which determines the relevance of input features for proper identification signals from the background, thereby allowing a dense deep-learning classifier to selectively focus on relevant features suppressing the irrelevant ones, consequently enhancing the accuracy and efficiency of signal recognition. We observed that a simple boosting algorithm with decision tree stumps trained over judiciously chosen jet-substructure-based features, nevertheless, exhibited higher efficiency. This indicates that our choice of input features made the feature space so fragmented that the dominance of a specific class (signal or different SM backgrounds) in a given subspace becomes prominent. 

The self-organizing maps, on the contrary, follows a different approach. The nodes, placed at the intersection of its grids, attempt to mimic the feature vector for each input data-point (event topologies) iteratively. The weights are adjusted so that the best matching unit (BMU) and the nodes in its neighborhood have their respective weights aligned with the matched feature vector. With training over each iteration (and updating weights assigned to each node), and subsequently over every batch (and subsequent updation of model parameters), the model captures the latent features specific to signal and different backgrounds. Finally, the grids end up forming clusters, with every node in each cluster possessing a weight vector resembling the underlying features that distinguish different event classes. As discussed in Section \ref{BSM}, each cluster predominantly captures events belonging to a specific Test dataset (\textit{see}, Tables \ref{tab:SR1} to \ref{tab:SR3}). The clusters formed on SOM grid are the planar projection of the feature space of the events, in a way that allows events with similar underlying topology captured in a single cluster.

Training SOM with SM top quark and QCD di-jet samples with assessment of topographic error in clustering, we observe that the emerged map fairly preserves the data topography within the grid. Further testing of the trained model with FCNC-driven top decay to a charm quark and a Higgs boson ($t \to cH$) demonstrates remarkable performance in the identification of anomalous regions for the training samples. We define three signal regions (SRs) based on the three-prong substructure of the candidate top jet and their kinematics. For SR-II, our model achieves a significance of $\sim 2\sigma$ for $\mathcal{BR}( t \to cH)$ = 0.05\%, establishing the performance of the model for segregating rare FCNC-mediated decays from overwhelming SM backgrounds. 
We emphasize that a possible explanation behind an enhanced performance of SOM over the existing approaches is embedded in the working principle of the algorithm, which learns to mimic the event features, leading to better capturing of patterns specific to the signal and different backgrounds in distinct clusters. {{Therefore, SOM has helped to gain insight from the data in an unsupervised manner, specifically by projecting different signal and background regions onto different, non-overlapping clusters within the Kohnen map. This in turn enables the identification of signal regions with high signal significances for probing the FCNC decay of the top quark. As shown in Section \ref{Result}, SOM has achieved fairly substantial significance despite the vanishingly small $t \to cH$ branching fraction on an overwhelming SM background. It is interesting to note that  
%We commenced with training the SOM architecture with SM backgrounds and used the trained model to probe the rare top-decay as anomalous events over the identified SM patterns. 
SOM can also be tailored for supervised learning, upon training the grids with the signal events and the SM backgrounds, thereby estimating the clusters and the event density in each cluster, using SOM as a model-dependent probe of the rare top-decay.}} 

SOMs, since its inception, has outgrown as tool for clustering and dimensionality reduction. As mentioned above, it has been adapted for supervised learning, enabling enhanced model performance while simultaneously preserving the topography of the training dataset. This versatility makes SOM a potential tool for BSM probes. 
Further improvements can be explored with alternative distance metrics, \textit{viz.}, the Mahalanobis distance, or the cosine similarity, in order to obtain more distinct clusters of signals and backgrounds in the Kohonen plane. Adaptive frameworks, such as Growing Self-Organizing Map (GSOM), which dynamically adapts its structure during training, by adding nodes as required based on data complexity, could further optimize the signal-to-background ratio. SOMs and its extensions are therefore, potential tools for future phenomenological explorations.
%==================================

\section{Acknowledgments}
AC acknowledges the financial support from the Department of Science and Technology, Government of India, under Grant No. IFA18-PH 224 (INSPIRE Faculty Award). The authors express their gratitude to the organizers of the ML4HEP-2023 conference held at ICTS Bangalore, India, for the initial development of this project. AC thanks Biplob Bhattacherjee for useful comments, and SC acknowledges Benjamin Fuks for insightful discussions at the Iwate Collider School-2025 at Iwate, Japan. The authors also thank Tousik Samui and Dharitree Bezboruah for their input in the initial stages of the work.

%=========================================================
\bibliographystyle{unsrt}
%\bibliography{reference}

\begin{thebibliography}{10}

\bibitem{Sakurai1960}
J.J. Sakurai.
\newblock Theory of strong interactions.
\newblock {\em Annals of Physics}, 11(1):1--48, 1960.

\bibitem{Scheck:1996ur}
F.~Scheck.
\newblock {\em {Electroweak and Strong Interactions. An Introduction to Theoretical Particle Physics}}.
\newblock Graduate Texts in Physics. Springer, 1996.

\bibitem{Pich:2007vu}
Antonio Pich.
\newblock {The Standard model of electroweak interactions}.
\newblock In {\em {2006 European School of High-Energy Physics}}, pages 1--49, 2007.

\bibitem{Pich:2005mk}
A.~Pich.
\newblock {The Standard model of electroweak interactions}.
\newblock In {\em {2004 European School of High-Energy Physics}}, pages 1--48, 2 2005.

\bibitem{Giudice2017}
Gian~F. Giudice.
\newblock The dawn of the post-naturalness era.
\newblock {\em Annual Review of Nuclear and Particle Science}, 67:429--447, 2017.

\bibitem{Sakharov1967}
A.~D. Sakharov.
\newblock Violation of cp invariance, c asymmetry, and baryon asymmetry of the universe.
\newblock {\em Journal of Experimental and Theoretical Physics Letters}, 5:24--27, 1967.

\bibitem{Peccei1977}
R.~D. Peccei and H.~R. Quinn.
\newblock Cp conservation in the presence of pseudoparticles.
\newblock {\em Physical Review Letters}, 38(25):1440--1443, 1977.

\bibitem{2010arXiv1005.1676L}
Joseph~D. {Lykken}.
\newblock {Beyond the Standard Model}.
\newblock {\em arXiv e-prints}, page arXiv:1005.1676, May 2010.

\bibitem{Apollinari:2015bam}
{High-Luminosity Large Hadron Collider (HL-LHC) : Preliminary Design Report}.
\newblock 12 2015.

\bibitem{Giovannozzi:2013pwa}
M.~Giovannozzi.
\newblock {The CERN LHC machine: Current status and future upgrade plans}.
\newblock {\em AIP Conf. Proc.}, 1560(1):686--690, 2013.

\bibitem{Larkoski:2017jix}
Andrew~J. Larkoski, Ian Moult, and Benjamin Nachman.
\newblock {Jet Substructure at the Large Hadron Collider: A Review of Recent Advances in Theory and Machine Learning}.
\newblock {\em Phys. Rept.}, 841:1--63, 2020.

\bibitem{Radovic:2018dip}
Alexander Radovic, Mike Williams, David Rousseau, Michael Kagan, Daniele Bonacorsi, Alexander Himmel, Adam Aurisano, Kazuhiro Terao, and Taritree Wongjirad.
\newblock {Machine learning at the energy and intensity frontiers of particle physics}.
\newblock {\em Nature}, 560(7716):41--48, 2018.

\bibitem{Plehn:2022ftl}
Tilman Plehn, Anja Butter, Barry Dillon, Theo Heimel, Claudius Krause, and Ramon Winterhalder.
\newblock {Modern Machine Learning for LHC Physicists}.
\newblock 11 2022.

\bibitem{MachineDAP}
Dimitri Bourilkov.
\newblock Machine and deep learning applications in particle physics.
\newblock {\em International Journal of Modern Physics A}, 34(35):1930019, 2019.

\bibitem{Feickert:2021ajf}
Matthew Feickert and Benjamin Nachman.
\newblock {A Living Review of Machine Learning for Particle Physics}.
\newblock 2 2021.

\bibitem{Kohonen_1982}
Teuvo Kohonen.
\newblock Self-organized formation of topologically correct feature maps.
\newblock {\em Biological Cybernetics}, 43(1):59--69, Jan 1982.

\bibitem{Kaplan:2008ie}
David~E. Kaplan, Keith Rehermann, Matthew~D. Schwartz, and Brock Tweedie.
\newblock {Top Tagging: A Method for Identifying Boosted Hadronically Decaying Top Quarks}.
\newblock {\em Phys. Rev. Lett.}, 101:142001, 2008.

\bibitem{Plehn:2010st}
Tilman Plehn, Michael Spannowsky, Michihisa Takeuchi, and Dirk Zerwas.
\newblock {Stop Reconstruction with Tagged Tops}.
\newblock {\em JHEP}, 10:078, 2010.

\bibitem{Butter:2017cot}
Anja Butter, Gregor Kasieczka, Tilman Plehn, and Michael Russell.
\newblock {Deep-learned Top Tagging with a Lorentz Layer}.
\newblock {\em SciPost Phys.}, 5(3):028, 2018.

\bibitem{Macaluso:2018tck}
Sebastian Macaluso and David Shih.
\newblock {Pulling Out All the Tops with Computer Vision and Deep Learning}.
\newblock {\em JHEP}, 10:121, 2018.

\bibitem{Chakraborty:2020yfc}
Amit Chakraborty, Sung~Hak Lim, Mihoko~M. Nojiri, and Michihisa Takeuchi.
\newblock {Neural Network-based Top Tagger with Two-Point Energy Correlations and Geometry of Soft Emissions}.
\newblock {\em JHEP}, 07:111, 2020.

\bibitem{Bhattacherjee:2022gjq}
Biplob Bhattacherjee, Camellia Bose, Amit Chakraborty, and Rhitaja Sengupta.
\newblock {Boosted top tagging and its interpretation using Shapley values}.
\newblock {\em Eur. Phys. J. Plus}, 139(12):1131, 2024.

\bibitem{Furuichi:2023vdx}
Amon Furuichi, Sung~Hak Lim, and Mihoko~M. Nojiri.
\newblock {Jet classification using high-level features from anatomy of top jets}.
\newblock {\em JHEP}, 07:146, 2024.

\bibitem{Brehmer:2024yqw}
Johann Brehmer, V\'\i{}ctor Bres\'o, Pim de~Haan, Tilman Plehn, Huilin Qu, Jonas Spinner, and Jesse Thaler.
\newblock {A Lorentz-Equivariant Transformer for All of the LHC}.
\newblock 11 2024.

\bibitem{Dong:2024xsg}
Zhongtian Dong, Dorival Gon\c{c}alves, Kyoungchul Kong, Andrew~J. Larkoski, and Alberto Navarro.
\newblock {Hadronic top quark polarimetry with ParticleNet}.
\newblock {\em Phys. Lett. B}, 862:139314, 2025.

\bibitem{Choi:2023slq}
Sang~Kwan Choi, Jinmian Li, Cong Zhang, and Rao Zhang.
\newblock {Automatic detection of boosted Higgs boson and top quark jets in an event image}.
\newblock {\em Phys. Rev. D}, 108(11):116002, 2023.

\bibitem{Collins:2018epr}
Jack~H. Collins, Kiel Howe, and Benjamin Nachman.
\newblock {Anomaly Detection for Resonant New Physics with Machine Learning}.
\newblock {\em Phys. Rev. Lett.}, 121(24):241803, 2018.

\bibitem{Amram:2020ykb}
Oz~Amram and Cristina~Mantilla Suarez.
\newblock {Tag N\textquoteright{} Train: a technique to train improved classifiers on unlabeled data}.
\newblock {\em JHEP}, 01:153, 2021.

\bibitem{Cheng:2020dal}
Taoli Cheng, Jean-Fran\c{c}ois Arguin, Julien Leissner-Martin, Jacinthe Pilette, and Tobias Golling.
\newblock {Variational autoencoders for anomalous jet tagging}.
\newblock {\em Phys. Rev. D}, 107(1):016002, 2023.

\bibitem{Pol:2020weg}
Adrian~Alan Pol, Victor Berger, Gianluca Cerminara, Cecile Germain, and Maurizio Pierini.
\newblock {Anomaly Detection With Conditional Variational Autoencoders}.
\newblock In {\em {Eighteenth International Conference on Machine Learning and Applications}}, 10 2020.

\bibitem{Finke:2021sdf}
Thorben Finke, Michael Kr\"amer, Alessandro Morandini, Alexander M\"uck, and Ivan Oleksiyuk.
\newblock {Autoencoders for unsupervised anomaly detection in high energy physics}.
\newblock {\em JHEP}, 06:161, 2021.

\bibitem{Ostdiek:2021bem}
Bryan Ostdiek.
\newblock {Deep Set Auto Encoders for Anomaly Detection in Particle Physics}.
\newblock {\em SciPost Phys.}, 12(1):045, 2022.

\bibitem{Ngairangbam:2021yma}
Vishal~S. Ngairangbam, Michael Spannowsky, and Michihisa Takeuchi.
\newblock {Anomaly detection in high-energy physics using a quantum autoencoder}.
\newblock {\em Phys. Rev. D}, 105(9):095004, 2022.

\bibitem{Alvi:2022fkk}
Sulaiman Alvi, Christian~W. Bauer, and Benjamin Nachman.
\newblock {Quantum anomaly detection for collider physics}.
\newblock {\em JHEP}, 02:220, 2023.

\bibitem{Dillon:2022tmm}
Barry~M. Dillon, Radha Mastandrea, and Benjamin Nachman.
\newblock {Self-supervised anomaly detection for new physics}.
\newblock {\em Phys. Rev. D}, 106(5):056005, 2022.

\bibitem{Golling:2023yjq}
Tobias Golling, Gregor Kasieczka, Claudius Krause, Radha Mastandrea, Benjamin Nachman, John~Andrew Raine, Debajyoti Sengupta, David Shih, and Manuel Sommerhalder.
\newblock {The interplay of machine learning-based resonant anomaly detection methods}.
\newblock {\em Eur. Phys. J. C}, 84(3):241, 2024.

\bibitem{Bickendorf:2023nej}
Gerrit Bickendorf, Manuel Drees, Gregor Kasieczka, Claudius Krause, and David Shih.
\newblock {Combining resonant and tail-based anomaly detection}.
\newblock {\em Phys. Rev. D}, 109(9):096031, 2024.

\bibitem{Metodiev:2023izu}
Eric~M. Metodiev, Jesse Thaler, and Raymond Wynne.
\newblock {Anomaly detection in collider physics via factorized observables}.
\newblock {\em Phys. Rev. D}, 110(5):055012, 2024.

\bibitem{Hammad:2024dsn}
A.~Hammad, Mihoko~M. Nojiri, and Masahito Yamazaki.
\newblock {Quantum similarity learning for anomaly detection}.
\newblock {\em JHEP}, 02:081, 2025.

\bibitem{10.21468/SciPostPhys.7.1.014}
Gregor Kasieczka, Tilman Plehn, Anja Butter, Kyle Cranmer, Dipsikha Debnath, Barry~M. Dillon, Malcolm Fairbairn, Darius~A. Faroughy, Wojtek Fedorko, Christophe Gay, Loukas Gouskos, Jernej~F. Kamenik, Patrick~T. Komiske, Simon Leiss, Alison Lister, Sebastian Macaluso, Eric~M. Metodiev, Liam Moore, Ben Nachman, Karl Nordström, Jannicke Pearkes, Huilin Qu, Yannik Rath, Marcel Rieger, David Shih, Jennifer~M. Thompson, and Sreedevi Varma.
\newblock {The Machine Learning landscape of top taggers}.
\newblock {\em SciPost Phys.}, 7:014, 2019.

\bibitem{Bose:2024pwc}
Camellia Bose, Amit Chakraborty, Shreecheta Chowdhury, and Saunak Dutta.
\newblock {Interplay of traditional methods and machine learning algorithms for tagging boosted objects}.
\newblock {\em Eur. Phys. J. ST}, 233(15-16):2531--2558, 2024.

\bibitem{Habermann:2023ksr}
Kai Habermann and Eckhard von Toerne.
\newblock {Self-Organizing Maps in High Energy Physics}.
\newblock {\em J. Phys. Conf. Ser.}, 2438(1):012120, 2023.

\bibitem{CMS:2024ubt}
Aram Hayrapetyan et~al.
\newblock {Search for flavor-changing neutral current interactions of the top quark mediated by a Higgs boson in proton-proton collisions at 13 TeV}.
\newblock 7 2024.

\bibitem{CMS:2021gfa}
Armen Tumasyan et~al.
\newblock {Search for flavor-changing neutral current interactions of the top quark and the Higgs boson decaying to a bottom quark-antiquark pair at $ \sqrt{s} $ = 13 TeV}.
\newblock {\em JHEP}, 02:169, 2022.

\bibitem{Chowdhury:2023jof}
Shreecheta Chowdhury, Amit Chakraborty, and Saunak Dutta.
\newblock {Boosted top tagging through flavour-violating interactions at the LHC}.
\newblock {\em Eur. Phys. J. C}, 85(3):231, 2025.

\bibitem{som:github}
Kai Habermann.
\newblock Code repository, 2021.
\newblock \url{https://github.com/KaiHabermann/SOM}.

\bibitem{Alwall:2011uj}
Johan Alwall, Michel Herquet, Fabio Maltoni, Olivier Mattelaer, and Tim Stelzer.
\newblock {MadGraph 5 : Going Beyond}.
\newblock {\em JHEP}, 06:128, 2011.

\bibitem{NNPDF:2014otw}
Richard~D. Ball et~al.
\newblock {Parton distributions for the LHC Run II}.
\newblock {\em JHEP}, 04:040, 2015.

\bibitem{Bierlich:2022pfr}
Christian Bierlich et~al.
\newblock {A comprehensive guide to the physics and usage of PYTHIA 8.3}.
\newblock 3 2022.

\bibitem{hepmc}
Matt Dobbs and Jorgen~Beck Hansen.
\newblock {The HepMC C++ Monte Carlo event record for High Energy Physics}.
\newblock {\em Comput. Phys. Commun.}, 134:41--46, 2001.

\bibitem{delphes}
J.~de~Favereau, C.~Delaere, P.~Demin, A.~Giammanco, V.~Lema\^\i{}tre, A.~Mertens, and M.~Selvaggi.
\newblock {DELPHES 3, A modular framework for fast simulation of a generic collider experiment}.
\newblock {\em JHEP}, 02:057, 2014.

\bibitem{Buchkremer:2013bha}
Mathieu Buchkremer, Giacomo Cacciapaglia, Aldo Deandrea, and Luca Panizzi.
\newblock {Model Independent Framework for Searches of Top Partners}.
\newblock {\em Nucl. Phys. B}, 876:376--417, 2013.

\bibitem{feynrules1}
Neil~D. Christensen and Claude Duhr.
\newblock {FeynRules - Feynman rules made easy}.
\newblock {\em Comput. Phys. Commun.}, 180:1614--1641, 2009.

\bibitem{feynrules2}
Adam Alloul, Neil~D. Christensen, C\'eline Degrande, Claude Duhr, and Benjamin Fuks.
\newblock {FeynRules 2.0 - A complete toolbox for tree-level phenomenology}.
\newblock {\em Comput. Phys. Commun.}, 185:2250--2300, 2014.

\bibitem{fastjet1}
Matteo Cacciari, Gavin~P. Salam, and Gregory Soyez.
\newblock {FastJet User Manual}.
\newblock {\em Eur. Phys. J. C}, 72:1896, 2012.

\bibitem{fastjet2}
Matteo Cacciari and Gavin~P. Salam.
\newblock {Dispelling the $N^{3}$ myth for the $k_t$ jet-finder}.
\newblock {\em Phys. Lett. B}, 641:57--61, 2006.

\bibitem{antikt}
Matteo Cacciari, Gavin~P. Salam, and Gregory Soyez.
\newblock {The anti-$k_t$ jet clustering algorithm}.
\newblock {\em JHEP}, 04:063, 2008.

\bibitem{Krohn:2009th}
David Krohn, Jesse Thaler, and Lian-Tao Wang.
\newblock {Jet Trimming}.
\newblock {\em JHEP}, 02:084, 2010.

\bibitem{ATLAS:2017bcq}
{Optimisation and performance studies of the ATLAS $b$-tagging algorithms for the 2017-18 LHC run}.
\newblock 2017.
\newblock All figures including auxiliary figures are available at https://atlas.web.cern.ch/Atlas/GROUPS/PHYSICS/PUBNOTES/ATL-PHYS-PUB-2017-013.

\bibitem{ATLAS:2015prs}
{Performance and Calibration of the JetFitterCharm Algorithm for c-Jet Identification}.
\newblock 1 2015.

\bibitem{NSubjettiness}
Jesse Thaler and Ken Van~Tilburg.
\newblock {Identifying Boosted Objects with N-subjettiness}.
\newblock {\em JHEP}, 03:015, 2011.

\bibitem{ROUSSEEUW198753}
Peter~J. Rousseeuw.
\newblock Silhouettes: A graphical aid to the interpretation and validation of cluster analysis.
\newblock {\em Journal of Computational and Applied Mathematics}, 20:53--65, 1987.

\bibitem{DBI}
David~L. Davies and Donald~W. Bouldin.
\newblock A cluster separation measure.
\newblock {\em IEEE Transactions on Pattern Analysis and Machine Intelligence}, PAMI-1(2):224--227, 1979.

\bibitem{scikit-learn}
F.~Pedregosa, G.~Varoquaux, A.~Gramfort, V.~Michel, B.~Thirion, O.~Grisel, M.~Blondel, P.~Prettenhofer, R.~Weiss, V.~Dubourg, J.~Vanderplas, A.~Passos, D.~Cournapeau, M.~Brucher, M.~Perrot, and E.~Duchesnay.
\newblock Scikit-learn: Machine learning in {P}ython.
\newblock {\em Journal of Machine Learning Research}, 12:2825--2830, 2011.

\bibitem{ATLAS:2018jqi}
Morad Aaboud et~al.
\newblock {Search for top-quark decays $t \to Hq$ with 36 fb$^{-1}$ of $pp$ collision data at $\sqrt{s}=13$ TeV with the ATLAS detector}.
\newblock {\em JHEP}, 05:123, 2019.

\bibitem{Cowan:2010js}
Glen Cowan, Kyle Cranmer, Eilam Gross, and Ofer Vitells.
\newblock {Asymptotic formulae for likelihood-based tests of new physics}.
\newblock {\em Eur. Phys. J. C}, 71:1554, 2011.
\newblock [Erratum: Eur.Phys.J.C 73, 2501 (2013)].

\end{thebibliography}

\end{document}